\documentclass{article}

\usepackage{graphicx,setspace}
\usepackage[margin=1.25in]{geometry}
\usepackage{caption,subfig}
\usepackage{amsmath,amssymb,amsthm}
\usepackage{caption}
\usepackage{tabularx,multirow}
\usepackage{appendix}

\renewcommand{\thefootnote}{\fnsymbol{footnote}}

\title{A production function with variable elasticity of substitution greater than one}

\author{Constantin Chilarescu \\
\it Laboratory CLERSE, University of Lille, France \\
\it E-mail: constantin.chilarescu@univ-lille.fr}
\date{}

\begin{document}

\maketitle
\renewcommand{\thefootnote}{\arabic{footnote}}

\begin{abstract}
The idea of this paper comes from the famous remark of Piketty and Zuckman: "It is natural to imagine that $\sigma$ was much less than one in the eighteenth and nineteenth centuries and became larger than one in the twentieth and twenty-first centuries. One expects a higher elasticity of substitution in high-tech economies where there are lots of alternative uses and forms for capital." The main aim of this paper is to prove the existence of a production function of variable elasticity of substitution with values greater than one.
\vspace{8 pt}


\noindent \textit{\\ JEL\ Classification:}: C20, C60, D20
\end{abstract}

\section{Introduction}
The elasticity of substitution is an essential macroeconomic concept and plays a crucial role in a multitude of theoretical economics topics. Due to its relevance, a long debate has emerged in the literature, not only for its value - lower or higher than one, but also on the two possible alternatives - constant elasticity or variable elasticity.
So far, there has been no consensus on these issues. More than that, some empirical studies confirm one of the hypotheses and others confirm exactly the opposite hypothesis, in particular that concerning its value in relation to one. As was pointed out by Grossman et al. $(2017)$, the size of the elasticity of substitution is much debated and still controversial. For example Chirinko $(2008)$, who surveyed and evaluated a large number of studies, suggests a value of the elasticity of substitution in the range of $0.4$ to $0.6$, Knoblach et al. $(2020)$ provide evidence that for the economy as a whole, the elasticity most likely falls in the range of $0.45$ to $0.87$. On the other hand, Karabarbounis and Neiman $(2014)$ estimate an elasticity of substitution greater than one, but Chirinko, Fazzari, and Meyer $(2011)$, Chirinko and Mallick $(2017)$ and Lawrence $(2015)$, all estimate elasticities below one.

The problem that the elasticity is constant or variable, is more or less, a theoretical approach. To decide between the two alternatives, we need a model, built on solid assumptions and then, if possible, analyzed via econometric methods to estimate its parameters.

One of the first empirical study on the elasticity of factor substitution, Arrow et al. $(1961)$ observed that factor substitutability can vary between industries and across countries. More recently Miyagiwa and Papageorgiou $(2007)$, concluded that the elasticity of factor substitution could be related to the level of economic development of a country. Duffy and Papageorgiou $(2000)$ use cross-country data to show that generally, richer countries have an elasticity of factor substitution above one, while poorer countries have an elasticity of factor substitution below one.

An interesting analysis on this issue is found in Piketty and Saez’s paper $(2014)$. They try to analyze the effect of the average annual real rate of return $(r)$ and the capital income ratio $\beta$, on the share of capital income in the national income, defined as $\alpha=r\beta$. In the standard economic model with perfectly competitive markets, $r$ is equal to the marginal product of capital. As the volume of capital $\beta$ rises, the marginal product $r$ tends to decline. The important question is whether $r$ falls more or less rapidly than the rise in $\beta$. If the standard hypothesis in economics are assumed, that is a unitary elasticity, then the fall in $r$, exactly offsets the rise in $\beta$, so that the capital share  is a technological constant. However, historical variations in capital shares are far from negligible. In recent decades, rich countries have experienced both a rise in $\beta$ and a rise in $\alpha$, which suggests that $\sigma$ is somewhat larger than one. This hypothesis has been suggested by Sato and Hoffman $(1968)$, who argued that as time passes and new technologies become available the opportunities for factor substitution are increased.

As noted above, there is a wide variety of estimates of the elasticity of substitution and this one can be interpreted as a missing specification of the production function and, allows us to consider other production functions that present a variable elasticity. Although there are a large number of papers devoted to production functions with a constant elasticity of substitution, the number of those which analysed production functions with variable elasticity is relatively limited and, unfortunately, written in the early $60$ and $70$. The first production function with the variable elasticity of substitution was proposed by Liu and Hildebrand $(1965)$ and a few years later by Lu $(1967)$. The elasticity of substitution of this function has an interesting property, being a decreasing function from a bounded upper limit, to one. Unfortunately, this property is true only for some $k$ greater than a given value $k_0 > 0$, and not for all $k >0$. Two other production functions were developed by Hoffman and Sato $(1968)$ and by Revankar $(1971)$. Unfortunately, for these latter functions, the elasticity of substitution increases infinitely when $k$ tends to infinity. The latest attempt to provide a production function with variable elasticity of substitution is provided by the paper of Chilarescu $(2019)$. The domain of variation of this elasticity of substitution is limited to the unit interval $(0, 1)$.

The main aim of this paper is to prove the existence of a production function with variable elasticity of substitution, but having the property that its elasticity is greater than one for all $k >0$. This result will be presented in the next section.

\section{A new production function with variable elasticity of substitution}
As in the paper of Sato and Hoffman $(1968)$, we consider the case of a production function assumed to be homogeneous of degree one and denote by: $y=f(k)=\frac{F(K,L)}{L}$, the output per-capita  and by $k=\frac{K}{L}$, the capital labor ratio.
Under this hypothesis, the marginal rate of substitution and the elasticity of substitution can be expressed in the following representation
\begin{equation}\label{eqmrs}
r(k)=\frac{y}{y^{\prime}}-k
\end{equation}
and
\begin{equation}\label{eqefs}
\sigma(k)=\frac{y^{\prime}\left(ky^{\prime}-y\right)}{kyy^{\prime\prime}},
\end{equation}
and thus they obtained the following result:
\begin{equation}\label{eqSH}
\frac{dy}{y}=\frac{dk}{k+\zeta\exp{\int{\frac{dk}{k \sigma(k)}}}},
\end{equation}
If we express the elasticity of substitution in terms of the marginal rate of substitution,
\begin{equation}\label{eqesmre}
\sigma(k)=\frac{r(k)}{kr^{\prime}(k)}=\frac{\frac{dk}{k}}{\frac{dr}{r}},
\end{equation}
then, the equation \eqref{eqSH} becomes
\begin{equation}\label{eqSH1}
\frac{dy}{y}=\frac{dk}{k+\zeta r(k)},\;\zeta>0.
\end{equation}
As it is well-known, the marginal rate of substitution is a positive increasing function. In order to obtain an explicit solution for the above differential equation, we assume that the function r(k) takes the following form:
\begin{equation}\label{eqdmrs}
r(k)= \frac{k\left[\alpha\left(1-\eta\right)k^{
\psi}+\beta\left(1-\mu\right)\right]}{\alpha\eta k^{\psi}+\beta\mu},\;\alpha,\;\beta,\;\eta,\;\mu\;\in (0, 1).
\end{equation}
Without loss of generality, we can suppose that $\zeta=1$ and thus the equation \eqref{eqSH1} will be written
\begin{equation}\label{eqSH2}
\frac{dy}{y}=\frac{dk}{k}\frac{\alpha\eta k^{\psi}
+\beta\mu}{\alpha k^{\psi}
+\beta}.
\end{equation}
If we denote $\mu=\theta$ and $\eta=\theta+\omega\psi$ then we obtain
\begin{equation}\label{eqSH3}
\frac{dy}{y}=\frac{dk}{k}\frac{\alpha\left[\theta+\omega\psi\right] k^{\psi}
+\beta\theta}{\alpha k^{\psi}
+\beta},
\end{equation}
whose solution is given by
\begin{equation}\label{eqsolpf}
y=f(k)=\gamma k^\theta\left[\alpha k^{\psi}+\beta\right]^{\omega},\; \gamma>0.
\end{equation}
It is just a simply exercise to prove that $y_0=f(0)=0$.
The first derivative is given by
\begin{equation}\label{eqderiv1}
f^{\prime}(k)=\gamma\frac{\alpha\left(\theta+\omega\psi
 \right)k^{\psi}+\beta\theta}{k^{1-\theta}\left(\alpha k^{\psi}+\beta
 \right)^{1-\omega}},
\end{equation}
and it is a positive function for all $k>0$.
$$\lim\limits_{k\rightarrow 0}f^{\prime}(k)=\gamma\lim\limits_{k\rightarrow 0}\frac{\alpha\left(\theta+\omega\psi
 \right)k^{\psi}+\beta\theta}{k^{1-\theta}\left(\alpha k^{\psi}+\beta
 \right)^{1-\omega}}=+\infty.$$
$$\lim\limits_{k\rightarrow +\infty}f^{\prime}(k)=
\gamma\lim\limits_{k\rightarrow \infty}\frac{\alpha\left(\theta+\omega\psi
 \right)k^{\psi}+\beta\theta}{k^{1-\theta}\left(\alpha k^{\psi}+\beta
 \right)^{1-\omega}}=0,\;\;\omega,\;\psi\in (0, 1),\;\;\theta+\omega\psi<1.$$
The second derivative is given by
\begin{equation}\label{eqderiv2}
f^{\prime\prime}=-f\frac{\alpha^{2}\left(\theta+\omega\psi\right)\left(1-\theta-\omega
\psi\right)k^{2\psi}+\alpha\beta
 \left[\psi\omega\left(1-\psi\right)+2\theta\left(1-\theta-\omega\psi\right)\right]k^\psi+\beta^
2\theta\left(1-\theta\right)}{{k}^{2}
 \left(\alpha{k}^{\psi}+\beta\right)^{2}},
\end{equation}
and it is negative for all $k>0$.
Substituting the results obtained from the equations \eqref{eqsolpf}, \eqref{eqderiv1} and \eqref{eqderiv2} into the equation  \eqref{eqefs} we get the following relation for the elasticity of substitution
\begin{equation}\label{eqes}
\sigma(k)=\frac{\left[\alpha\left(\theta+\omega\psi\right)k^{\psi}
+\beta\theta\right]\left[\alpha\left(1-\theta-\omega\psi\right)k^{\psi}+\beta\left(1-\theta\right)\right]}
{\left[\alpha\left(\theta+\omega\psi\right)k^{\psi}
+\beta\theta\right]\left[\alpha\left(1-\theta-\omega\psi\right)k^{\psi}
+\beta\left(1-\theta\right)\right]-\alpha\beta\omega\psi^2k^\psi}\geq 1.
\end{equation}
Passing to the limit for $k\rightarrow 0$ and $k\rightarrow +\infty$, we obtain
$$\lim\limits_{k\rightarrow 0}\sigma(k) = 1,\;\mbox{and}
\;\lim\limits_{k\rightarrow \infty}\sigma(k) =\left\{\begin{array}{cc}
                                                       1 &\;\; if \;\;2\psi \geq 1 \\\\
                                                       1+\frac{\omega\psi^2}{2\theta(1-\theta-\omega\psi)
                                                       +\omega\psi(1-\psi)} & \;\;if\;\; 2\psi < 1
                                                     \end{array}\right.
$$
The first derivative of the elasticity of substitution is given by
\begin{equation}\label{eqfdes}
\sigma^{\prime}(k)=\frac{\alpha\beta\omega\psi^3 k^{\psi-1}\left[\beta^2\theta\left(1-\theta\right)-\alpha^2 \left(\theta+\omega\psi\right)\left(1-\theta-\omega\psi\right)k^{2\psi}
\right]}{\left\{\left[\alpha\left(\theta+\omega\psi\right)k^{\psi}
+\beta\theta\right]\left[\alpha\left(1-\theta-\omega\psi\right)k^{\psi}
+\beta\left(1-\theta\right)\right]-\alpha\beta\omega\psi^2k^\psi\right\}^2}.
\end{equation}
If $$k\in \left(0,\left[\frac{\beta^2\theta\left(1-\theta\right)}{\alpha^2 \left(\theta+\omega\psi\right)\left(1-\theta-\omega\psi\right)}\right]^{\frac{1}{2\psi}}\right),$$
then $\sigma^{\prime}(k)$ is positive and therefore $\sigma(k)$ will be an increasing function. For all $$k>\left[\frac{\beta^2\theta\left(1-\theta\right)}{\alpha^2 \left(\theta+\omega\psi\right)\left(1-\theta-\omega\psi\right)}\right]^{\frac{1}{2\psi}}$$
$\sigma^{\prime}(k)$ is negative and therefore $\sigma(k)$ will be a decreasing function.

The relative share of capital $(\pi_k)$ and the relative share of labor $(\pi_l)$, are given by
\begin{equation}\label{eqmpcl}
\pi_k=\frac{kr}{f}=\frac{\alpha\left(\theta+\omega\psi\right)k^{\psi}
+\beta\theta}{\left(\alpha k^{\psi}+\beta\right)},\;\;\pi_l=1-\frac{kr}{f}=1-\frac{\alpha\left(\theta+\omega\psi\right)k^{\psi}+\beta\theta
}{\alpha k^{\psi}+\beta}.
\end{equation}
It is just a simple exercise to prove that the relative share of capital is a bounded increasing function, the relative share of labor is a bounded decreasing function and,
$$\lim\limits_{k\rightarrow\infty}\pi_k=\theta+\omega\psi<1\;\;\mbox{and}\;\;\lim\limits_{k
\rightarrow\infty}\pi_l=1-\theta-\omega\psi>0.$$
\section{Properties, numerical simulations and conclusions}
The new production function has the following properties:
\begin{enumerate}
  \item [a.] $f(k)\geq 0$, $f^{\prime}(k)\geq 0$, $f^{\prime\prime}(k)\leq 0$ for all $k\geq 0$, $\lim\limits_{k\rightarrow 0}f^{\prime}(k)=+\infty$ and $\lim\limits_{k\rightarrow +\infty}f^{\prime}(k)=0$,
  \item [b.] the elasticity of substitution is, first an increasing function up to a bounded upper limit and then a decreasing function,
  \item [c.] if $\omega=0$, then the new production function reduces to the Cobb-Douglas function,
  \item [d.] if $\theta=0$ and $\omega=\frac{1}{\psi}$, then the new production function reduces to the $CES$ function.
\end{enumerate}
Let us consider the following benchmark values for the economy:
$$\theta=0.6,\;\;\omega=0.5,\;\;\psi=0.7,\;\;\alpha=0.2,\;\;\beta=0.8,\;\;\gamma=1.05$$
The results of the numerical simulation procedure are presented in the following graphs.
\begin{center}
\includegraphics[width=6cm]{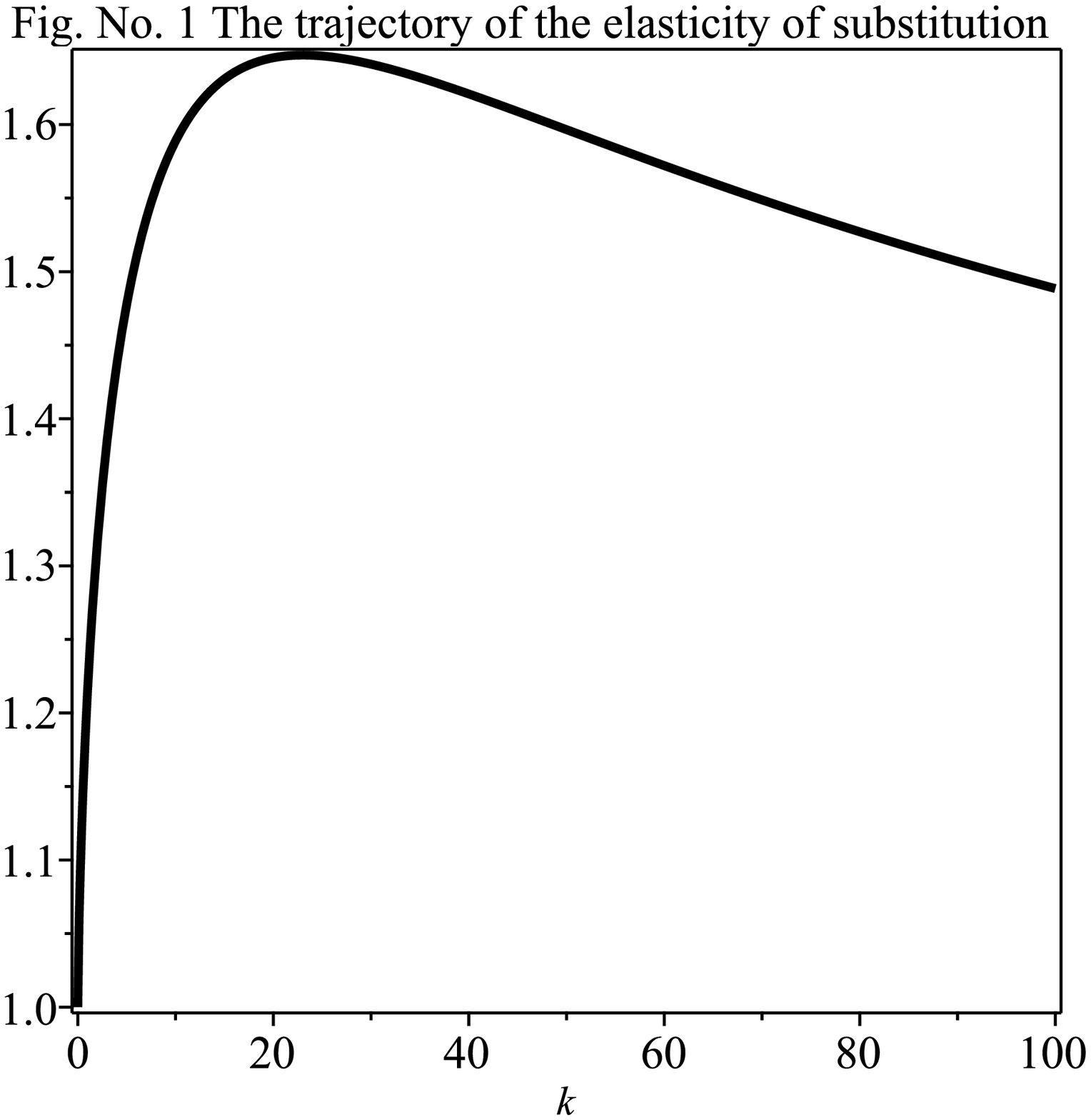}\;\;\;\;\;\includegraphics[width=6cm]{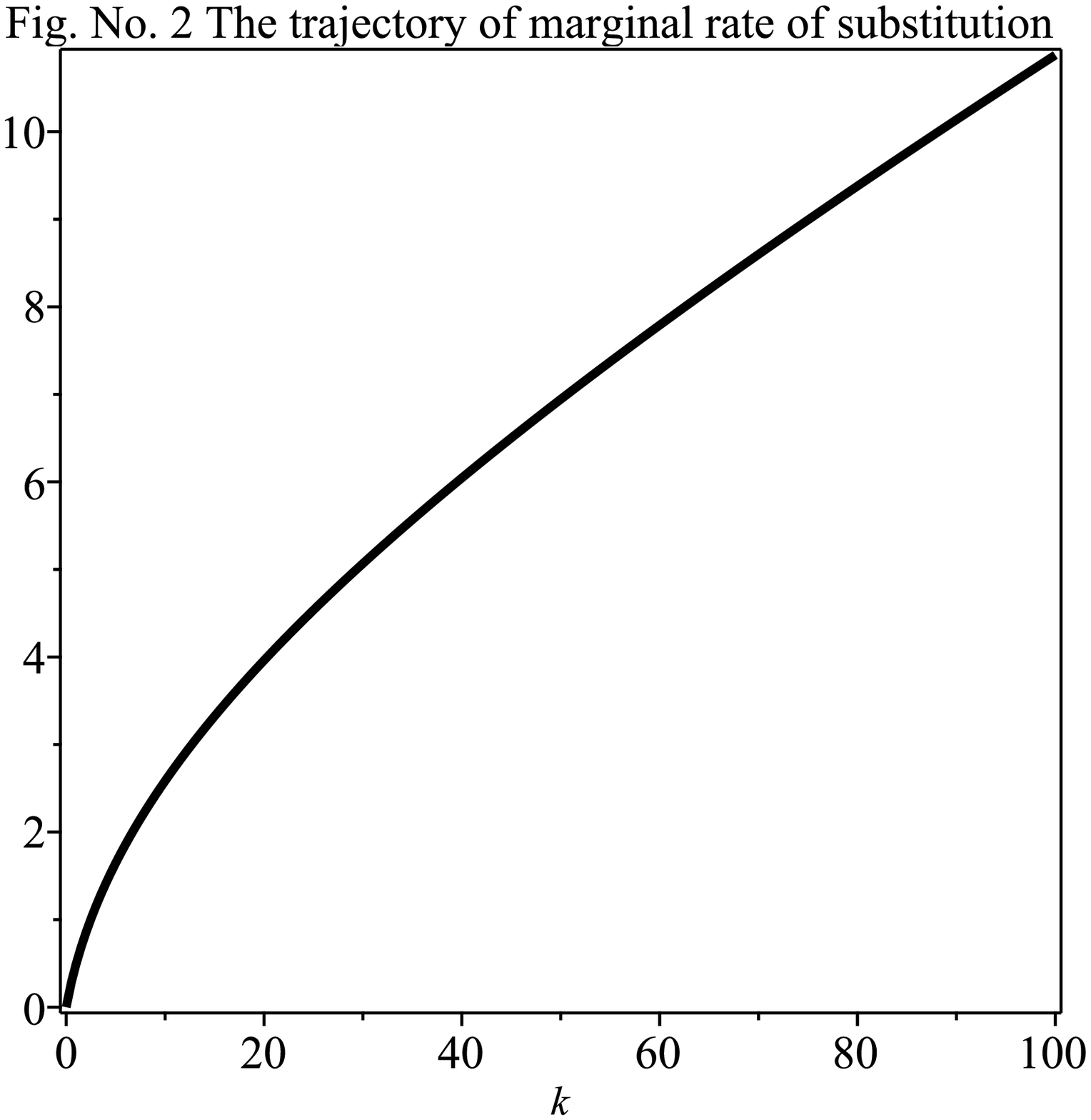}
\end{center}
\begin{center}
\includegraphics[width=6cm]{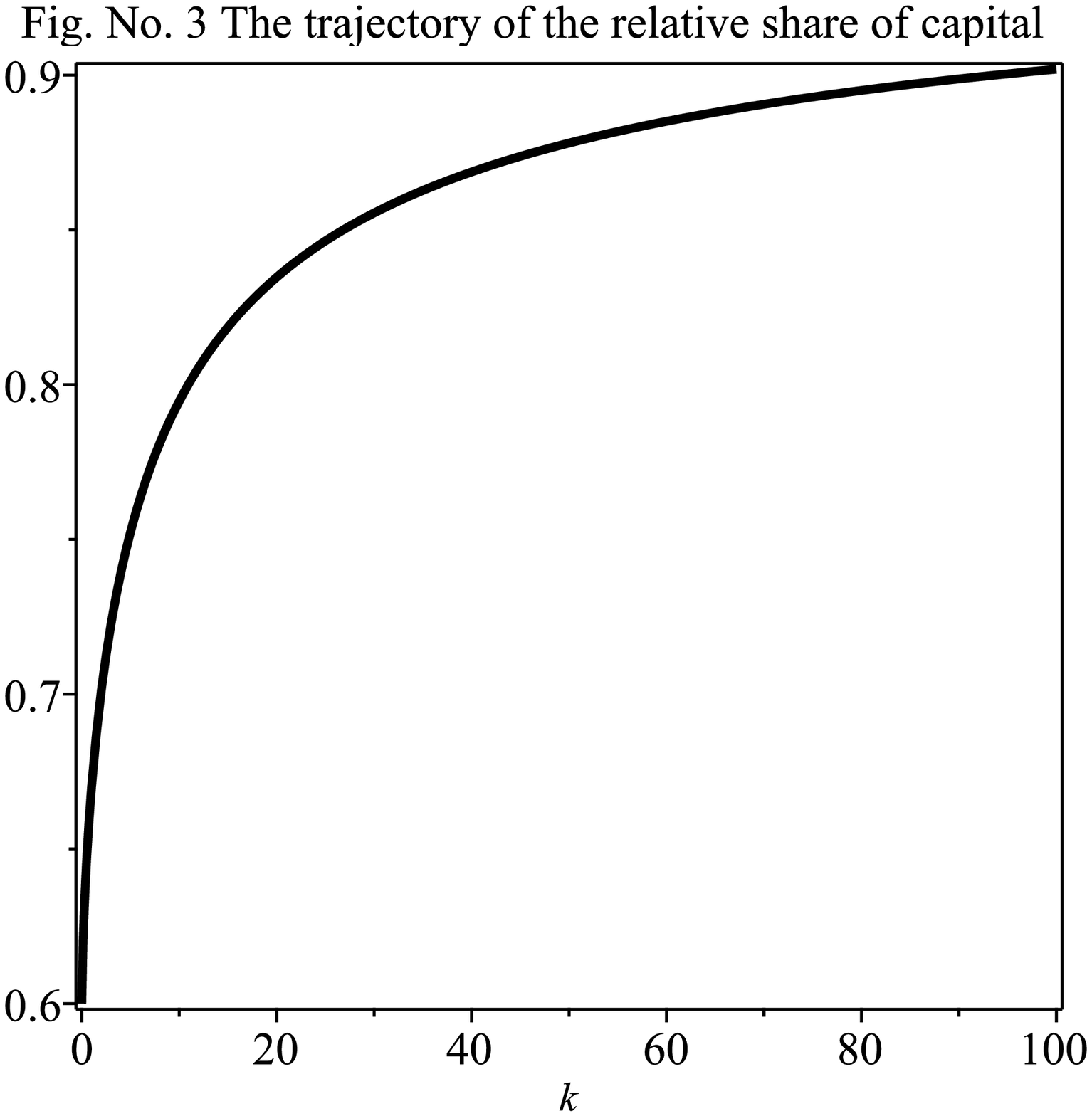}\;\;\;\;\;\includegraphics[width=6cm]{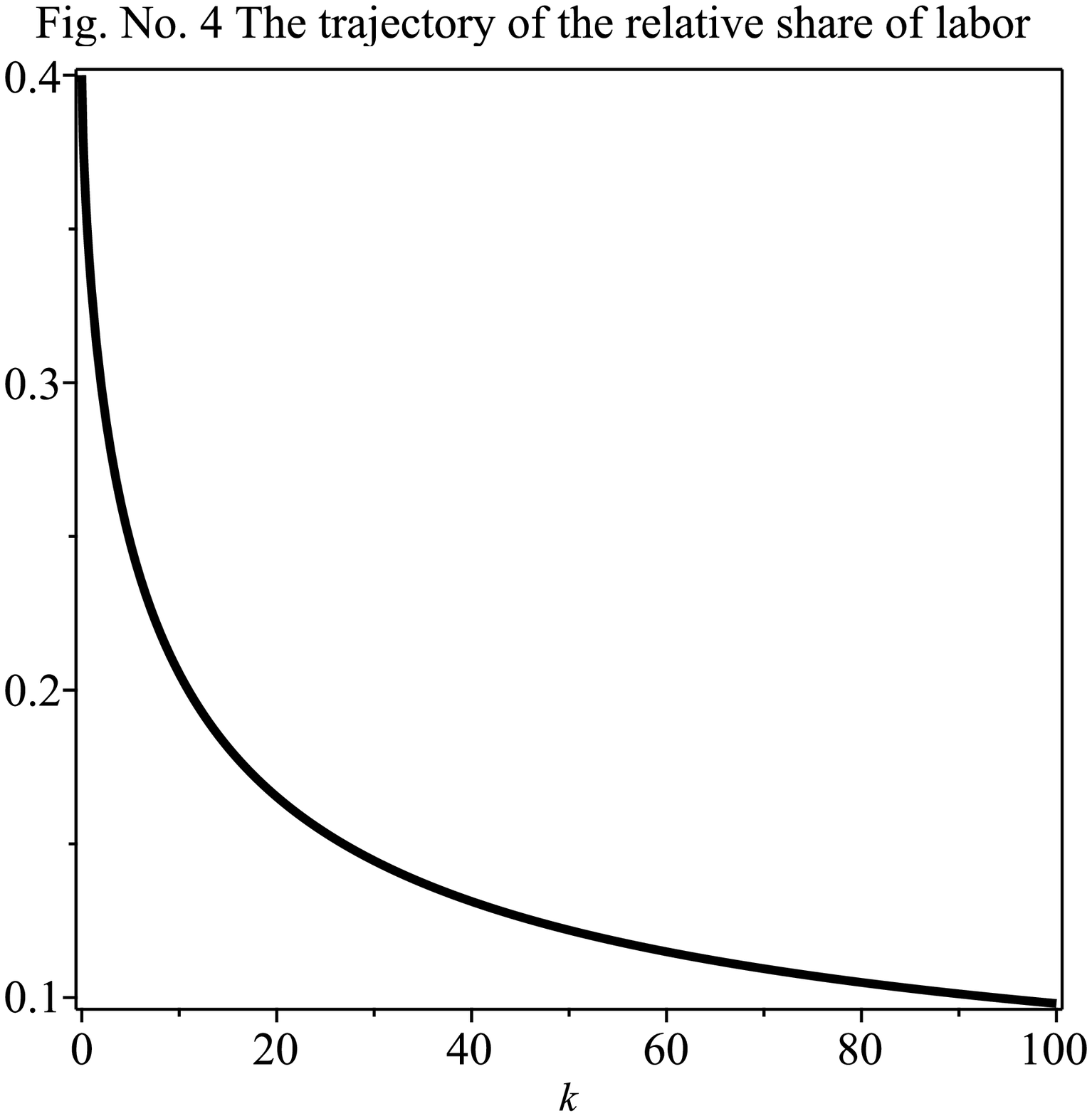}
\end{center}
We can observe that the elasticity of substitution increases with $k$ to an upper bounded limit and then slowly decreases to the bounded inferior limit.


\begin{thebibliography}{99}
\bibitem{AMCS}  Arrow K. J., Chenery H. B., Minhas B. S. and Solow R. M. $1961$. Capital-Labor Substitution and Economic Efficiency. The Review of Economics and Statistics, $43$ $(3)$, $225 - 250$.
\bibitem{CHI} Chilarescu C. $2019$.    A Production Function with Variable Elasticity of Factor Substitution. Economics Bulletin, $39$ $(4)$, $2343 - 2360$.
\bibitem{CK} Chirinko R. S. $2008$. Sigma: The long and short of it. Journal of Macroeconomics, $30$, $671 - 686$.
\bibitem{CF} Chirinko R. S., Fazzari S. M. and Meyer A. P. $2011$. A New Approach to Estimating Production Function Parameters. The Elusive Capital-Labor Substitution Elasticity.
    Journal of Business and Economic Statistics, $29$ $(4)$, $587 - 594$.
\bibitem{CM} Chirinko R. S., and Mallick D., $2017$. The Substitution Elasticity, Factor Shares, and the Low-Frequency Panel Model. American Economic Journal: Macroeconomics, $9$ $(4)$, $225 - 253$.
\bibitem{DP} Duffy J. and Papageorgiu C., $2000$. A Cross-Country Empirical Investigation of the
Aggregate Production Function Specification. Journal of Economic Growth, $5$, $87 - 120$.
\bibitem{GH} Grossman G. M., Helpman E., Oberfield E. and Sampson T. $2017$. Balanced Growth Despite Uzawa,
American Economic Review, Vol. $107$, $1293 - 1312$.
\bibitem{KN} Karabarbounis L. and Neiman B., $2014$. The Global Decline of the Labor Share. The Quarterly Journal of Economics, $61 - 103$.
\bibitem{KR} Knoblach M., Roessler M. and Zwerschke P. $2020$. The Elasticity of Substitution Between Capital and Labour in the US Economy: A Meta‐Regression Analysis. Oxford Bulletin of Economics and Statistics, Vol. $82$ $(1)$, $62 - 82$.
\bibitem{LA} Lawrence R. Z. $2015$. Recent Declines in Labor’s Share in US Income: A Preliminary Neoclassical Account. National Bureau of Economic Research Working Paper $21296$.
\bibitem{LH} Liu T. C. and Hildebrand H., $1965$. Manufacturing Production Functions in the United States $1957$. Ithaca: Cornell University Press, $1965$.
\bibitem{LUD} Lu Yaoji, $1967$. Variable elasticity of substitution production functions, technical change and factor shares. Retrospective Theses and Dissertations. $http://lib.dr.iastate.edu/rtd/3406$
\bibitem{LF}  Lu Y. and Fletcher R. F., $1968$. A Generalization of the CES Production Function. The Review of Economics and Statistics, $50$ $(4)$, $449 - 452$.
\bibitem{Mal}  Mallick D., $2012$. The role of the elasticity of substitution in economic growth:
A cross-country investigation. Labour Economics, $19$, $682 - 694$.
\bibitem{MP} Miyagiwa K. and Papageorgiou C. $2007$. Endogenous aggregate elasticity of substitution. Journal of Economic Dynamics and Control, $31$ $(9)$, $2899 - 2919$.
\bibitem{PS} Piketty T. and Saez E. $2014$. Inequality in the Long Run.” Science $344$, $838 - 843$.
\bibitem{PZ} Piketty T. and Zuckman G. $2014$. Capital is Back: Wealth-Income
Ratios in Rich Countries $1700 - 2010$. The Quarterly Journal of Economics, $129$ $(3)$,
$1255 - 1310$.
\bibitem{RK} Revankar N. S., $1971$. A Class of Variable Elasticity of Substitution Production Functions.
Econometrica, $39$$(1)$, $61 - 71$.
\bibitem{SH}  Sato R. and Hoffman R. F., $1968$. Production Functions with Variable Elasticity of Factor Substitution: Some Analysis and Testing. The Review of Economics and Statistics, $50$ $(4)$, $453 - 460$.
\end{thebibliography}
\end{document}